\documentclass[letterpaper, 10 pt, conference]{ieeeconf}  % Comment this line out if you need a4paper

\usepackage{cite}
\usepackage{graphicx}
\usepackage{textcomp}
\usepackage{balance} % To balance the last page columns
\usepackage{textcomp}
\def\BibTeX{{\rm B\kern-.05em{\sc i\kern-.025em b}\kern-.08em
    T\kern-.1667em\lower.7ex\hbox{E}\kern-.125emX}}
\usepackage{multirow}
\usepackage{url}
\usepackage{bm}
\usepackage{color}
\usepackage[LY1]{fontenc}
\usepackage{comment}
\usepackage{array}
\newcolumntype{P}[1]{>{\centering\arraybackslash}p{#1}}
\newcolumntype{M}[1]{>{\centering\arraybackslash}m{#1}}
\usepackage{subcaption}
\usepackage{afterpage}

\usepackage{amssymb,amsfonts}
\usepackage[fleqn]{amsmath}
\usepackage{threeparttable}
\usepackage{siunitx}
\usepackage{soul}
\usepackage{booktabs}
\usepackage{algorithm}
\usepackage{algcompatible}

\usepackage{colortbl}

\definecolor{Gray}{gray}{0.85}
\definecolor{LightCyan}{rgb}{0.88,1,1}
\definecolor{LightGreen}{rgb}{0,0.88,0}
\definecolor{LightRed}{rgb}{0.88,0,0}

\IEEEoverridecommandlockouts                              % This command is only needed if 
\title{\LARGE \bf
Neonatal Face and Facial Landmark Detection from Video Recordings
}

\author{Ethan Grooby$^{1,2,3}$, \IEEEmembership{Student Member, IEEE},
Chiranjibi Sitaula$^{1}$,
Soodeh Ahani$^{2,3}$,
Liisa Holsti$^{2}$,
Atul Malhotra$^{4}$,
\\Guy A. Dumont$^{2,3}$, \IEEEmembership{Life Fellow, IEEE}, and
Faezeh Marzbanrad$^{1}$, \IEEEmembership{Senior Member, IEEE}
\thanks{*E. Grooby  acknowledges the support of the MIME-Monash Partners-CSIRO sponsored PhD research support program and Research Training Program (RTP). A. Malhotra's research is supported by the Kathleen Tinsley Trust and a Cerebral Palsy Alliance Research Grant. F. Marzbanrad is supported by the Veski Victoria Fellowship 2021. The study is supported by the Monash Institute of Medical Engineering (MIME) and Monash Data Future Institute (MDFI).}
\thanks{$^{1}$Department of Electrical and Computer Systems Engineering, Monash University, Melbourne, VIC, Australia.}
\thanks{$^{2}$BC Children's Hospital Research Institute, Vancouver, BC, Canada.}   
\thanks{$^{3}$Department of Electrical and Computer Engineering, University British Columbia, Vancouver, BC, Canada.}  
\thanks{$^{4}$Monash Newborn, Monash Children’s Hospital and Department of Paediatrics, Monash University, Melbourne, VIC, Australia.}
\thanks{email: \tt\small ethan.grooby@monash.edu}}

\begin{document}

\maketitle
\thispagestyle{empty}
\pagestyle{empty}

%%%%%%%%%%%%%%%%%%%%%%%%%%%%%%%%%%%%%%%%%%%%%%%%%%%%%%%%%%%%%%%%%%%%
\begin{abstract}
This paper explores automated face and facial landmark detection of neonates, which is an important first step in many video-based neonatal health applications, such as vital sign estimation, pain assessment, sleep-wake classification, and jaundice detection. Utilising three publicly available datasets of neonates in the clinical environment, 366 images (258 subjects) and 89 (66 subjects) were annotated for training and testing, respectively. Transfer learning was applied to two YOLO-based models, with input training images augmented with random horizontal flipping, photo-metric colour distortion, translation and scaling during each training epoch. Additionally, the re-orientation of input images and fusion of trained deep learning models was explored. Our proposed model based on YOLOv7Face outperformed existing methods with a mean average precision of 84.8\% for face detection, and a normalised mean error of 0.072 for facial landmark detection. Overall, this will assist in the development of fully automated neonatal health assessment algorithms. 

%\newline

\indent \textit{Clinical relevance}— Accurate face and facial landmark detection provides an automated and non-contact option to assist in video-based neonatal health applications. 
\end{abstract}

%%%%%%%%%%%%%%%%%%%%%%%%%%%%%%%%%%%%%%%%%%%%%%%%%%%%%%%%%%%%%%%%%%%%%%%%%%%%%%%%
\section{INTRODUCTION}
Video monitoring provides a non-contact method for monitoring neonates in clinical, home and remote environments. Typically, the detection of the patient's facial area as a region of interest (ROI) with associated facial landmarks is an essential first step. After this step, ROI tracking between video frames and subsequent analysis of the tracked ROI would then be performed. 

Video monitoring using face ROIs has received increasing interest for infant cardio-respiratory monitoring recently~\cite{dosso2022nicuface,khanam2021non,kyrollos2021noncontact}.  Cardio-respiratory monitoring is crucial for early detection, prediction of prognosis, and continued monitoring of major complications during the neonatal period~\cite{grooby2023artificial}. It assists clinicians in providing timely and appropriate care to possibly minimise morbidity and mortality~\cite{grooby2023artificial}. 

Similarly, video face monitoring has been investigated for the timely detection of neonates in pain~\cite{brahnam2020neonatal}. This monitoring allows for treatment to prevent long-term effects on neurological and behavioural development~\cite{brahnam2020neonatal}. Additionally, facial video monitoring for the detection of jaundice~\cite{dosso2022nicuface} and sleep-wake classification~\cite{awais2021hybrid} has been investigated in newborns.   

Face and facial landmark detection for neonates, especially in clinical environments, has proven difficult~\cite{wan2022infanface, huang2021neonatal, hausmann2022robust, dosso2022nicuface}. Neonates have less well-defined facial features compared to adults~\cite{wan2022infanface}, poses/postures are more varied, and common obstructions, such as bottle feeding and pacifiers are prevalent~\cite{wan2022infanface, huang2021neonatal, hausmann2022robust}. Additionally, in the clinical environment, background scenes are complicated, medical devices and interventions may be involved, and varying lighting conditions pose additional challenges~\cite{dosso2022nicuface, hausmann2022robust}. Additionally, many face and facial landmark detectors have been trained on predominantly adult data~\cite{wan2022infanface,dosso2022nicuface}. Overall, this has led to poor performance in neonatal face ROI detection, which results in manual ROI selection of the neonatal face being common~\cite{dosso2022nicuface,paul2020non,gibson2019non}. %aarts2013non

This paper presents an overview of several existing adult and neonatal face/landmark detectors in Section~\ref{sec:existing}. Improved deep learning models using transfer learning, image re-orientation and fusion techniques are  proposed in Section~\ref{sec:methods}. Existing and proposed methods are then evaluated and discussed in Sections~\ref{sec:results} and \ref{sec:discussion}.

\section{EXISTING WORKS}
\label{sec:existing}
\subsection{Adult Face and Facial Landmark Detection}
\label{sec:adult}
Current state-of-the-art methods for face and facial landmark detection have utilised machine learning, especially deep learning \cite{kumar2019face, liu2016ssd,Deng2020CVPR,bazarevsky2019blazeface}. A traditional method for adult face detection is OpenCV's Haar cascade method~\cite{opencv_library}. In this method, Haar features are extracted and an AdaBoost algorithm is used to detect the face~\cite{opencv_library}. 

For deep learning methods, Dlib face detection uses a histogram of oriented gradients and a convolutional neural network (CNN) based on max-margin for face detection~\cite{dlib09}. With multi-task cascaded CNN (MTCNN)~\cite{zhang2016joint}, the correlation between the detection and alignment of the face and associated landmarks was utilised to boost performance. MTCNN was broken up into three stages to predict the face and five landmarks (\textit{i.e.}, right eye, left eye, nose, right mouth, and left mouth) on a course to a fine level, and trained on the WIDER FACE dataset~\cite{yang2016wider}. 

More recently, single-shot multibox detectors (SSD) have been popular for face detection~\cite{liu2016ssd,Deng2020CVPR,bazarevsky2019blazeface}. SSD~\cite{liu2016ssd} utilises a set of default bounding boxes over different aspect ratios and scales per feature map location. During face detection, each default box generates an output of if a face is present, and appropriate adjustments to the box to match the face~\cite{liu2016ssd}. 

RetinaFace~\cite{Deng2020CVPR} is a novel SSD that unified face bounding box prediction and 2D facial landmark detection. The model was trained on the WIDER FACE dataset~\cite{yang2016wider}, and outputs five facial landmarks (\textit{i.e.} right and left eye, nose, right mouth, and left mouth). % ResNet-50 model backbone

BlazeFace~\cite{bazarevsky2019blazeface} is a lightweight face detector tailored for mobile GPU inference with models for front-facing and rear-facing cameras. The feature extraction network is inspired by MobileNetV1/V2, and the anchor scheme is modified from SSD~\cite{bazarevsky2019blazeface}. MediaPipe provides an updated implementation of the BlazeFace architecture and outputs face detection and 6 facial landmarks (\textit{i.e.} right and left eyes, nose, centre mouth, and left and right ears)~\cite{bazarevsky2019blazeface}. 

% SSD can process 9.20 frames per second whereas fps rates are 6.50 for Haar cascade, 1.57 for Dlib Hog, 1.54 for MTCNN. In other words, SSD is the fastest one. 

\subsection{Neonatal Face and Facial Landmarks Detection}
\label{sec:neonatal}
Similarly to adult-based existing work, the majority of past works in the neonatal space have focused on deep learning-based methods \cite{kyrollos2021noncontact,khanam2021non,dosso2022nicuface, hausmann2022robust}. In particular, the YOLO framework has been especially popular \cite{khanam2021non,dosso2022nicuface, hausmann2022robust}. 

For face detection, Awais \textit{et al.}~\cite{awais2021hybrid} proposed a semi-automated intensity-based method. RGB colour space was converted into the CIELAB colour space. Then, an appropriate threshold is then determined by examining intensity values within each CIELAB channels~\cite{awais2021hybrid}. Finally, the facial region is determined as the region with the highest number of linked portions~\cite{awais2021hybrid}.  

Olmi \textit{et al.}~\cite{olmi2022aggregate} developed a fully automated face detection method for newborns in the neonatal intensive care unit (NICU) from the aggregate channel feature algorithm. The model was trained on newborn data collected at the Neuro-physiopathology and Neonatology Clinical Units of AOU Careggi, Firenze, Italy~\cite{olmi2022aggregate}. 

For deep learning-based face detectors, Khanam \textit{et al.}~\cite{khanam2021non} utilised YOLOv3 model~\cite{redmon2018yolov3}, pre-trained on the MS COCO dataset~\cite{lin2014microsoft}. Transfer learning was then applied using 473 images of neonates obtained from the internet~\cite{lin2014microsoft}. 

Kyrollos \textit{et al.}~\cite{kyrollos2021noncontact} constructed a deep learning model with a ResNet backbone and RetinaNet head~\cite{lin2017focal}, pre-trained on the ImageNet dataset~\cite{deng2009imagenet}. This model was  further trained on patients admitted to the NICU of the Children's Hospital of Eastern Ontario~\cite{dosso2022nicuface}. 

Building on~\cite{kyrollos2021noncontact}, Dosso \textit{et al.}~\cite{dosso2022nicuface} developed NICUface. NICUface is designed specifically for neonatal face detection in complex NICU environments. Two base models, YOLOv5Face~\cite{qi2021yolo5face} and RetinaFace~\cite{Deng2020CVPR} were explored for NICUface development, as these models performed the best in initial testing. These models were then subsequently trained on publicly available datasets; the newborn baby heart rate estimation database (NBHR)~\cite{huang2021neonatal} and iCOPEvid dataset~\cite{brahnam2020neonatal}, as well as data previously collected from the Children's Hospital of Eastern Ontario~\cite{dosso2022nicuface}. Overall, NICUface with YOLOv5Face base performed the best compared to existing base models and NICUface with RetinaFace base. NICUface is also able to output 5 facial landmarks (\textit{i.e.} right, and left eyes, nose, right mouth and left mouth); however, the updated model was not re-trained specifically for landmark detection and so only qualitative assessment was performed. 

Similarly, Hausmann \textit{et al.}~\cite{hausmann2022robust} developed a face detector based on YOLOv5 model~\cite{glenn_jocher_2020_4154370}. The model was trained on images taken from Tampa General Hospital with Male 47.3\%: Female 52.7\% sex breakdown, and an ethnicity breakdown of 72.7\% Caucasian, 20\% African-American, and 7.3\% Asian. Overall, the model in~\cite{hausmann2022robust} did significantly better than YOLO model~\cite{glenn_jocher_2020_4154370} baselines.

For facial landmarks specifically, Wan \textit{et al.}~\cite{wan2022infanface} developed a model from  HRNet's facial landmark detection~\cite{wang2020deep}. The model requires the face detection bounding box  as its input, and then outputs 68 facial landmarks relating to eyes, eyebrows, nose, mouth and the outline of the face~\cite{wan2022infanface}. This model~\cite{wan2022infanface} was further trained on infant images obtained from the internet. 

%Non-contact physiological monitoring of preterm infants in the Neonatal Intensive Care Unit villarroel2019non
%Using our proposed multi-task Convolutional Neural Network (CNN), time periods during which the infant was present or absent from the incubator were automatically detected from the video recordings. Regions of interest (ROI) corresponding to the skin were segmented from each video frame.
%Patient detection and skin segmentation
%- Proposed CNN network performed the joint task of image classification and segmentation
%- Multi-task network has a shared core network, implemented using the VGG-16 architecture. 

% Pain Assessment From Facial Expression: Neonatal Convolutional Neural Network (N-CNN)
% We applied ZFace [37] face tracker in each frame of the recorded videos to detect the face and obtain 49 facial landmark points.

\section{METHODS}
\label{sec:methods}
\subsection{Data Acquisition}

\subsubsection{Datasets}
Three Publicly available datasets~\cite{brahnam2020neonatal},~\cite{wan2022infanface}, and \cite{huang2021neonatal} are used for this study.

NBHR~\cite{huang2021neonatal} consists of 1,130 videos of 257 neonates at 0-6 days old, totalling 9.6 hours of recordings with synchronous photoplethysmogram and heart rate physiological signals. The newborn infants were recruited from the Department of Obstetrics and Gynaecology, Xinzhou People’s Hospital, China. Biological sex was approximately equal (Male 48.6\%: Female 51.4\%). Camera angle relative to each baby's location varied, facial occlusion was present in a subset of videos from bottle feeding and sleep position, and a variety of natural hospital room illuminations were obtained~\cite{huang2021neonatal}. 

Infant Annotated Faces (InfAnFace)~\cite{wan2022infanface}, contains 410 images of infants obtained from Google, YouTube and advertisements. These images were annotated with 68 facial landmarks and pose attributes~\cite{wan2022infanface}. % what about ethnicities?

iCOPEvid dataset~\cite{brahnam2020neonatal} contains  234 videos (20 seconds/video) of 49 neonates experiencing a set of noxious stimuli, a period of rest and an acute pain stimulus. Neonatal videos were taken in the neonatal nursery at Mercy Hospital in Springfield Missouri, USA, from infants aged between 34 to 70 hours~\cite{brahnam2020neonatal}. Biological sex was approximately balanced (Male 53.1\%: Female 46.9\%), and 16 neonates were swaddled compared to 33 free to move~\cite{brahnam2020neonatal}. Ethnicity was distributed as 41 Caucasian, 1 African-American, 1 Korean, 2 Hispanic, and 4 interracial~\cite{brahnam2020neonatal}.

\subsubsection{Data extraction}
For NBHR~\cite{huang2021neonatal} and iCOPEvid~\cite{brahnam2020neonatal} datasets, the first frame was extracted from each video to be annotated. Additionally, neonatal images of the same subject that were similar between videos were removed to provide a diverse dataset. From the InfAnFace dataset~\cite{wan2022infanface}, images were excluded if not representative of the hospital environment or neonatal age period. 

Overall, 455 images from 324 neonates were obtained for annotation. These images consisted of 352 images from 257 neonates, 18 images from 18 infants, and 85 images from 49 neonates from the NBHR~\cite{huang2021neonatal}, InfAnFace~\cite{wan2022infanface}, and iCOPEvid~\cite{brahnam2020neonatal} datasets, respectively. 

\subsubsection{Annotations}
We manually annotated the provided dataset. The face bounding box captured the area from the forehead to the chin and between the ears. In cases of partial occlusions, if subsections of the face were still visible/partially visible in the area of occlusion, they were annotated as the face. Otherwise, if a complete region of the face was occluded, it was not included in the bounding box, as the size of the face occluded was difficult to infer. Additionally, 6 facial landmarks were identified, namely; right eye, left eye, nose, right mouth, centre mouth and left mouth. 
% might add a figure here of facial annotation

\subsection{Transfer Learning}
\label{sec:transfer}
\subsubsection{Base Models}
Based on the performance of face detection reported in table~\ref{tab1:face_detect}, YOLOv5 and YOLOv7Face~\cite{qi2021yolo5face} were chosen as base models. YOLOv5 starting weights were obtained from~\cite{hausmann2022robust}. YOLOv7Face starting weights were the 'yolov7' model on GitHub \url{https://github.com/derronqi/yolov7-face}, which was pre-trained on the WIDER FACE dataset~\cite{yang2016wider}. 

\subsubsection{Data Augmentation}
During each training epoch, random horizontal flip (50\% probability), photo-metric colour distortion (\textit{i.e.} $\pm 1\%$, $\pm 10\%$, $\pm 10\%$  HSV Hue, Saturation and Value), translation (\textit{i.e.} $\pm 10\%$), and scaling (\textit{i.e.} $\pm 10\%$) augmentation were performed. 

\subsubsection{Hyperparameters}
For YOLOv5, hyperparameters were set similarly as in~\cite{dosso2022nicuface}. An initial learning rate of 0.0032, final learning rate of 0.12, warm-up momentum of 0.5, final momentum of 0.843, warm-up period of 2 epochs, intersection over union (IoU) threshold of 20\%, and optimised using stochastic gradient descent were used during training. 

For YOLOv7Face, hyperparameters were set similarly to the sample settings in YOLOv7Face~\cite{qi2021yolo5face}. Learning rate of 0.01, warm-up momentum of 0.8, final momentum of 0.937, warm-up period of 3 epochs, IoU threshold of 20\%, and optimised using stochastic gradient descent were used during training.

\subsubsection{Training} 
The data was split into approximately 80\% training and 20\% test, resulting in 366 images from 258 subjects for training, and 89 images from 66 subjects for test evaluation. It was ensured that all images from the same subject were in only the training or the test sets. 

Within the training set, data was further split for 5-fold cross-validation, while ensuring all subjects' images were within the same fold. The number of epochs to train the model and the number of layers to freeze within the models were then explored. 

Based on initial cross-validation results, the final YOLOv5 model was trained over 100 epochs with no layers frozen, and the final YOLOv7Face model was trained over 50 epochs with no layers frozen. Both models were trained with a batch size of 16. 

\subsection{Orientation} 
\label{sec:orientation}
Existing detectors have been typically trained on images where the face is upright \cite{dosso2022nicuface}. Since neonatal poses are variable, rotating the image into four different scenarios could improve the likelihood of a more optimal pose for the model.  

Hence, utilising the same face and facial landmark detector, four images were inputted into the detector. These four images were the 0, 90, 180, and 270 degrees rotated versions of the original image. The image that outputted the highest confidence detection was then used for face and landmark estimation. 

\subsection{Fusion}
\label{sec:fusion}
Based on the results presented in Tables~\ref{tab1:face_detect} and ~\ref{tab1:landmark_detect}, two fusion methods were considered. 

Firstly, as RetinaNet with re-orientation and the proposed YOLOv7Face produced the lowest mean normalised error (MNE) for facial landmarks, a fusion model based on majority voting was considered. Whereby, the method that has the highest confidence (\textit{i.e.} max confidence in four RetinaFace outputs compared to the proposed YOLOv7Face), was used for facial landmarks estimation. With this fusion model, the average of execution time is approximately 2.09\,s per image.

To reduce the computational time, MediaPipe front camera model was used to estimate the optimal orientation of the input image for RetinaFace. In this fusion model, the average execution time is reduced to approximately 1.58\,s per image. 

%Combining several face detectors to obtain higher accuracy bounding box estimator was explored using two methods \textcolor{red}{ add methods with their references(\textit{i.e.} ...., and ....)}. 

%\subsubsection{Majority Voting}
%Among the selected face detectors, the final bounding box for the face is determined as the detected face by using the method that outputted the highest level of confidence. 

%\subsubsection{Overlap}
%Among the selected face detectors, the top face detectors within $\pm 10\%$ confidence level are chosen. The final bounding box for the face is the overlap region between these selected face detectors. 

\subsection{Evaluation}
For face and facial landmark detection, as there is only one neonate present in each image, the confidence threshold was set to 5\% to maximise the number of detections, with the highest confidence detection chosen for evaluation. If no face is detected, this was defined as a false negative. 

\subsubsection{Face Detection}
\label{sec:detect}
Average precision (AP) is defined by \eqref{eq:precision}, with a particular IoU threshold. IoU is defined as the area of overlap divided by the area of union between the reference and estimated bounding box. IoU threshold represents when to classify the face detection as a true positive (TP) or a false positive (FP). Two AP metrics are considered, average precision for IoU threshold of 50\% (AP50), and mean average precision for IoU thresholds 50\% to 95\% in 5\% increments (mAP). 

%Average recall (AR) is defined by \eqref{eq:recall}, with similar IoU thresholds to get AR50 and mAR. 

\begin{equation}
\label{eq:precision}
\begin{split}
Precision = \frac{TP}{TP+FP}
\end{split}
\end{equation}

%\begin{equation}
%\label{eq:recall}
%\begin{split}
%Recall = \frac{TP}{TP+FN}
%\end{split}
%\end{equation}

\subsubsection{Facial Landmark Detection}
\label{sec:landmark}
MNE was used for the assessment of landmark detection between reference (ref) and estimated (est) landmarks for all test samples (N). Where the normalisation distance was the reference face bounding box area (ref\_bbox) \eqref{eq:norm_error},\eqref{eq:mne}.

\begin{equation}
\label{eq:norm_error}
\begin{split}
norm\_error(i) & =  \frac{\sqrt{(ref^{i}_x-est^{i}_x)^2+(ref^{i}_y-est^{i}_y)^2}}{\sqrt{ref\_bbox^{i}_{width} \times ref\_bbox^{i}_{height}}}\\
\end{split}
\end{equation}

\begin{equation}
\label{eq:mne}
\begin{split}
MNE &= \frac{\sum_{i=1}^{N} norm\_error(i)}{N}\\
\end{split}
\end{equation}

\begin{table}[th]
    \begin{center}
\caption{Comparative Study of Face Detectors}
    \begin{tabular}{ |M{2.2cm}|M{1.0cm}|M{1.0cm}|M{1.0cm}|M{1.0cm}|}
    \hline \textbf{Method}
    & \textbf{AP50 (\%)}
    & \textbf{mAP (\%)}
    & \textbf{Empty (\%)}
    & \textbf{Time (ms)}
    \\
    \hline
    \multicolumn{5}{|c|}{\textbf{Adult Existing Methods} (Section~\ref{sec:adult})} 
    \\
    \hline
    MediaPipe~\cite{bazarevsky2019blazeface}\newline Front 
    & 86.0
    & 30.5
    & 3.4
    & \textbf{9}
    \\
    \hline  
    MediaPipe~\cite{bazarevsky2019blazeface}\newline Back 
    & 41.2
    & 16.5
    & 80.9
    & 11
    \\
    \hline  
    OpenCV~\cite{opencv_library}
    & 10.0
    & 7.0
    & 88.8
    & 115
    \\
    \hline  
    SSD~\cite{liu2016ssd}
    & 75.0
    & 55.0
    & 95.5
    & 123
    \\
    \hline  
    Dlib~\cite{dlib09}
    & \textbf{100.0}
    & 50.0
    & 97.8
    & 126
    \\
    \hline  
    MTCNN~\cite{zhang2016joint}
    & 85.0
    & 51.5
    & 77.5
    & 686
    \\
    \hline  
    RetinaFace~\cite{Deng2020CVPR}
    & 97.1
    & 55.9
    & 21.3
    & 182
    \\
    \hline  
    YOLOv7Face~\cite{qi2021yolo5face}
    & \textbf{100.0}
    & 64.6
    & \textbf{0.0}
    & 1364
    \\
    \hline  
    \multicolumn{5}{|c|}{\textbf{Neonatal Existing Methods} (Section~\ref{sec:neonatal})} 
    \\
    \hline
    NICUface~\cite{dosso2022nicuface}
    & 83.1
    & 61.5
    & 33.7
    & 1096
    \\
    \hline 
    YOLOv5~\cite{hausmann2022robust}
    & 87.7
    & 46.4
    & 9.0
    & 211
    \\
    \hline 
    \multicolumn{5}{|c|}{\textbf{Proposed Methods} (Section~\ref{sec:transfer})} 
    \\
    \hline
    YOLOv5*
    & 98.8
    & 81.8
    & 6.7
    & 211
    \\
    \hline
    YOLOv7Face*
    & \textbf{100.0}
    & \textbf{84.7}
    & \textbf{0.0}
    & 1364
    \\
    \hline
    \multicolumn{5}{|c|}{\textbf{Orientation} (Section~\ref{sec:orientation})} 
    \\
    \hline
    MediaPipe \newline Front 
    & 88.8
    & 30.8
    & \textbf{0.0}
    & 35
    \\
    \hline
    RetinaFace 
    & \textbf{100.0}
    & 67.5
    & 2.2
    & 728
    \\
    \hline
    NICUface
    & \textbf{100.0}
    & 79.0
    & \textbf{0.0}
    & 4383
    \\
    \hline 
    YOLOv5 
    & 87.6
    & 47.5
    & \textbf{0.0}
    & 845
    \\
    \hline 
    YOLOv7Face
    & 98.9
    & 72.6
    & \textbf{0.0}
    & 5457
    \\
    \hline  
    YOLOv5*  
    & 98.9
    & 80.0
    & 2.2
    & 845
    \\
    \hline
    YOLOv7Face*
    & \textbf{100.0}
    & 83.4
    & \textbf{0.0}
    & 5457
    \\
    \hline
        \multicolumn{5}{p{0.46\textwidth}}{*Proposed re-trained models. \newline
        AP50 and mAP are defined in Section~\ref{sec:detect}, and Time refers to mean computational time as defined in Section~\ref{sec:time}. Empty refers to the percentage of images with no face detected.}
    \end{tabular}
    \label{tab1:face_detect}
\end{center}    
\end{table}

\begin{table}[th]
    \begin{center}
\caption{Comparative Study of Facial Landmark Detectors}
    \begin{tabular}{ |M{2.16cm}|M{0.7cm}|M{0.7cm}|M{0.7cm}|M{0.8cm}|M{0.8cm}|}
    \hline \textbf{Method}
    & \textbf{All}
    & \textbf{Eyes}
    & \textbf{Nose}
    & \textbf{Mouth}
    & \textbf{Empty (\%)}
    \\
    \hline
    \multicolumn{6}{|c|}{\textbf{Adult Existing Methods} (Section~\ref{sec:adult})} 
    \\
    \hline
    MediaPipe~\cite{bazarevsky2019blazeface}\newline Front 
    & 0.192
    & 0.204
    & 0.184
    & 0.175
    & 3.4
    \\
    \hline   
    RetinaFace~\cite{Deng2020CVPR}
    & 0.095
    & 0.102
    & 0.096
    & 0.088
    & 21.3
    \\
    \hline  
    YOLOv7Face~\cite{qi2021yolo5face}
    & 0.228
    & 0.242
    & 0.227
    & 0.215
    & \textbf{0.0}
    \\
    \hline  
    \multicolumn{6}{|c|}{\textbf{Neonatal Existing Method} (Section~\ref{sec:neonatal})} 
    \\
    \hline
    NICUface~\cite{dosso2022nicuface}
    & 0.320
    & 0.323
    & 0.267
    & 0.343
    & 33.7
    \\
    \hline 
    \multicolumn{6}{|c|}{\textbf{Proposed Method} (Section~\ref{sec:transfer})} 
    \\
    \hline
    YOLOv7Face*
    & 0.072
    & 0.077
    & 0.062
    & 0.072
    & \textbf{0.0}
    \\
    \hline
    \multicolumn{6}{|c|}{\textbf{Orientation} (Section~\ref{sec:orientation})} 
    \\
    \hline
    MediaPipe\newline Front 
    & 0.109
    & 0.119
    & 0.102
    & 0.095
    & \textbf{0.0}
    \\
    \hline   
    RetinaFace
    & 0.075
    & \textbf{0.073}
    & 0.073
    & 0.078
    & 2.2
    \\
    \hline  
    YOLOv7Face
    & 0.190
    & 0.198
    & 0.180
    & 0.186
    & \textbf{0.0}
    \\
    \hline 
    NICUface
    & 0.083
    & 0.074
    & \textbf{0.047}
    & 0.109
    & \textbf{0.0}
    \\
    \hline 
    YOLOv7Face*
    & 0.111
    & 0.122
    & 0.087
    & 0.112
    & \textbf{0.0}
    \\
    \hline
    \multicolumn{6}{|c|}{\textbf{Fusion Methods} (Section~\ref{sec:fusion})} 
    \\
    \hline		
    RetinaFace + \newline YOLOv7Face*
    & \textbf{0.070}
    & 0.075
    & 0.059
    & 0.070
    & \textbf{0.0}
    \\
    \hline
    MediaPipe Front + \newline
    RetinaFace + \newline 
    YOLOv7Face*
    & 0.071
    & 0.078
    & 0.062
    & \textbf{0.069}
    & \textbf{0.0}
    \\
    \hline
    \multicolumn{6}{p{0.44\textwidth}}{*Proposed re-trained models. \newline
    Facial landmark detection is assessed using MNE as defined in Section~\ref{sec:landmark}. 'All' refers to the MNE of all facial landmarks combined, whereas 'Eyes', 'Nose' and 'Mouth' refer to MNE for particular landmarks. Empty refers to the percentage of images with no facial landmark detection, which was excluded in the calculation of MNE.}
    \end{tabular}
    \label{tab1:landmark_detect}
\end{center}    
\end{table} 

\subsubsection{Execution Time} 
\label{sec:time}
We evaluated the average execution time per image during the inference/test phase as another assessment metric. A faster method is more suitable for real-time processing and mobile-phone implementation. The average execution time per image was calculated using Python on a MacBook Pro CPU 2.3\,GHz 8-Core Intel~i9.

\subsubsection{Significance Testing} Paired Wilcoxon signed-rank test was used to determine if IoU and normalised errors were significantly better, with p-value threshold of 0.05 used. 

\section{RESULTS}
\label{sec:results}
Table~\ref{tab1:face_detect} presents the face detection results. Within the existing methods, OpenCV, SSD, Dlib and MTCNN in particular struggled to detect the neonate's face with more than three-quarters of faces not detected. Out of the existing methods, YOLOv7Face performed significantly the best with AP50 of 100\% and mAP of 64.6\%. Both proposed methods showed significant improvements compared to existing methods, with the re-trained YOLOv7Face model performing the best with AP50 of 100\% and mAP of 84.7\%. 

Re-orienting the image for the highest confidence showed improved results in all methods with the exception of the proposed YOLOv7Face model. In particular, existing methods RetinaFace, YOLOv5, YOLOv7Face, and NICUface showed significant improvements. Even with re-orientation, the proposed models still performed significantly better, with the exception of the proposed YOLOv5 model and the re-orientated YOLOv5 model compared NICUface, where the p-values were 0.3 and 0.1, respectively.

Table~\ref{tab1:landmark_detect} presents the facial landmark detection results. Similarly with face detection, re-orientation significantly improved the results for existing methods, whereas the proposed YOLOv7Face model performed worse. Compared with all existing and re-oriented existing methods, the proposed YOLOv7Face model shows significant improvement in overall MNE. The proposed YOLOv7Face model produced MNE values of 0.072, 0.077, 0.062, and 0.072 for all, eyes, nose, and mouth landmarks respectively. 

Compared to using just the proposed YOLOv7Face model, fusion model 1 (RetinaFace + YOLOv7Face*) produced insignificant changes. Whilst this fusion model produced on average lower errors, 25 landmark estimates were worse compared to 20 that improved. Whereas fusion model 2 (MediaPipe Front + RetinaFace + YOLOv7Face*) produced overall significant but minor improvements, with MNE of 0.071, 0.078, 0.062, and 0.069 for all, eyes, nose, and mouth landmarks respectively.

\section{DISCUSSION}
\label{sec:discussion}
Overall, the proposed models showed improvement in both face and facial landmark detection. 
One note of caution regarding the results is that the NICUface model was used unaltered from \cite{dosso2022nicuface}. However, this model was trained on the same iCOPEvid and NBHR datasets. Hence, the results presented for NICUface may be an overstatement. %and why this model was not considered in the proposed fusion models despite promising results. 

One limitation of these models is the average execution time of 211\,ms and 1,364\,ms for YOLOv5 and YOLOv7Face, respectively. With execution times being relatively long compared to other methods, this may hinder real-time processing. However, face detection is typically only required in the first step prior to ROI tracking which is computationally less expensive. Therefore real-time processing may still be possible in the overall implementation of a neonatal video monitoring system with these models. Future work looking into the utilisation of BlazeFace \cite{bazarevsky2019blazeface} for real-time mobile phone processing may be useful. 

The re-orientation method proved to be a highly effective process for existing methods to both increase the success rate of detecting the face and facial landmarks, as well as improve their accuracy. This highlights a simple process to address the various poses/postures that the neonate may be in. However, this would require four times the execution time to perform this process. 

For the proposed models, YOLOv5 produced insignificant improvements and YOLOv7Face had deteriorated results using the re-orientation method. A potential explanation for this is these models were trained on raw input images that encompassed the various neonatal poses and postures without re-orientation. Instead, the re-orientation changed the relationship of the newborn with regard to the clinical environment. 

\section{CONCLUSIONS}
This paper presented two re-trained YOLO-based models that showed significant improvement in face and facial landmark detection. Additionally, the re-orientation of images showed marked improvement in existing methods, whereas the proposed fusion of methods showed minor improvements. Future work into neonatal face segmentation to better localise the face should be investigated. 


%%%%%%%%%%%%%%%%%%%%%%%%%%%%%%%%%%%%%%%%%%%%%%%%%%%%%%%%%%%%%%%%%%%%%%%%%%%%%%%%

%%%%%%%%%%%%%%%%%%%%%%%%%%%%%%%%%%%%%%%%%%%%%%%%%%%%%%%%%%%%%%%%%%%%%%%%%%%%%%%%

%%%%%%%%%%%%%%%%%%%%%%%%%%%%%%%%%%%%%%%%%%%%%%%%%%%%%%%%%%%%%%%%%%%%%%%%%%%%%%%%
%\section*{APPENDIX}

%\section*{ACKNOWLEDGMENT}

%%%%%%%%%%%%%%%%%%%%%%%%%%%%%%%%%%%%%%%%%%%%%%%%%%%%%%%%%%%%%%%%%%%%%%%%%%%%%%%%

\bibliographystyle{IEEEtran}
\bibliography{IEEEabrv,references}

\end{document}